\documentclass[a4paper,10pt]{article}
\usepackage{inputenc}
\usepackage[normalem]{ulem}
\usepackage{latexsym,amssymb,amsmath,amssymb,xspace,theorem}
\usepackage{multicol}
\usepackage{t1enc}

\newcommand{\Ajoute}[1]{{\uwave{#1}}} 
\newcommand{\Enleve}[1]{{\xout{#1}} }

\theoremstyle{plain} \theoremheaderfont{\scshape}

\newtheorem{Thm}{\bf Theorem}
\newtheorem{Lem}[Thm]{\bf Lemma}
\newtheorem{Clm}{Claim}[Thm]

\newtheorem{Nots}[Thm]{\bf Notations}

\newtheorem{Conj}[Thm]{{\bf Conjecture}}

\newtheorem{Cor}[Thm]{ \bf Corollary}
{\theorembodyfont{\rmfamily}
 
 \newtheorem{Rem}[Thm]{\bf Remark}

}

\newenvironment{Prf}{{\bf \noindent Proof } }{\hfill$\square$\\}
\newenvironment{PrfClaim}{{\bf Proof }}{{\hfill\tiny{$\blacksquare$\\}}}

\newcommand{\ignore}[1]{}
\newcommand{\ligne}{\vspace{0.5cm}}

\title{Reed's Conjecture on hole expansions}
\author{J.-L. Fouquet\\
 L.I.F.O., Facult\'e des Sciences, B.P. 6759 \\
 Universit\'e d'Orl\'eans, 45067 Orl\'eans Cedex 2, FR \thanks{email~:jean-luc.fouquet@univ-orleans.fr}
\and J.-M. Vanherpe \\
L.I.F.O., Facult\'e des Sciences, B.P. 6759 \\
Universit\'e d'Orl\'eans, 45067 Orl\'eans Cedex 2, FR\thanks{email~:jean-marie.vanherpe@univ-orleans.fr}
}

\begin{document}

\maketitle

\begin{abstract}
In 1998, Reed conjectured that for any graph $G$, $\chi(G) \leq \lceil \frac{\omega(G) + \Delta(G)+1}{2}\rceil$, 
where $\chi(G)$, $\omega(G)$, and $\Delta(G)$ respectively denote the chromatic number, the clique number and the maximum degree of $G$. 
In this paper, we study this conjecture for some {\em expansions} of graphs, that is graphs obtained  with the well known operation {\em composition} of graphs.\\
We prove that Reed's Conjecture holds for expansions of bipartite graphs,  for expansions of odd holes  where the minimum chromatic number of the components is even, 
when some component of the expansion has chromatic number $1$ or when a component induces a bipartite graph. Moreover, Reed's Conjecture holds if all components have the same chromatic number,
if the components have chromatic number at most $4$ and when the odd hole has length $5$. Finally, when $G$ is an odd hole expansion,  we prove $\chi(G)\leq\lceil\frac{\omega(G)+\Delta(G)+1}{2}\rceil+1$.
\end{abstract}

\section{Introduction}
We consider here simple and undirected graphs. For terms which are not defined we refer to Bondy and Murty \cite{BonMur08}. 

The chromatic number of a graph $G$, denoted by $\chi(G)$, is the minimum number of colors required to a proper colouring of the graph, that is to colour the vertices of $G$ so that no two adjacent vertices 
receive the same colour~; the size of the largest clique (independent set) in $G$ is called the {\em clique number} ({\em independence number}) of $G$, and denoted by $\omega(G)$ ($\alpha(G)$)~; 
the maximum degree of $G$, denoted $\Delta(G)$ is the maximum number of neighbours of a vertex over all vertices of $G$.

Bounding the chromatic number of a graph in terms of others graphs parameters attracted much attention in the past. For example, it is well know that for any graph $G$ we have 
$\omega(G)\leq\chi(G)\leq\Delta(G)+1$. This upper bound was reduced to $\Delta(G)$ by Brooks \cite{Bro41} in 1941 for connected graphs which are not complete graphs neither odd cycles.

In 1998 Reed \cite{Ree98} stated the following Conjecture also known as {\em Reed's Conjecture}: 
\begin{Conj}\cite{Ree98}\label{Conj:Reed}
 For any graph $G$, $\chi(G)\leq\lceil\frac{\omega(G)+\Delta(G)+1}{2}\rceil$.
\end{Conj}

This conjecture has been stated true for some restrictions of the graph parameters. Hence Conjecture \ref{Conj:Reed} holds true when $\chi(G)>\lceil \frac{|V(G)|}{2}\rceil$ 
(see \cite{Rab08}),  when $\chi(G)\leq\omega(G)+2$ \cite{GerRab08}, when $\alpha(G)=2$ \cite{Kin09,KohSch10} or when $\Delta(G)\geq|V(G)|-\alpha(G)-4$ (see \cite{KohSch10}).

Some classes of graphs also verify Conjecture \ref{Conj:Reed}. That's trivially the case for perfect graphs (a graph $G$ is said to be perfect if $\chi(H)=\omega(H)$ for every induced subgraph $H$ of $G$), for graphs with disconnected complement \cite{Rab08}
for almost split graphs (an almost-split graph is a graph that can be partitioned into a maximum independent set and a graph having independence number at most $2$) or particular classes of triangle free graphs 
\cite{KohSch10} and for classes defined with forbidden configurations such that $(2K_2, C_4)$-free graphs, odd hole free graphs \cite{AraKarSub11} or some particular classe of
 $P_5$-free graphs \cite{AraKarSub11}.

\ligne The well known operation {\em composition} of graphs, also called {\em expansion} in \cite{AraKarSub11} is defined as follows~:

Given a graph $H$ on $n$ vertices $v_0\ldots v_{n-1}$ and a familly of graphs $G_0\ldots G_{n-1}$, an {\em expansion} of $H$, denoted $H(G_0\ldots G_{n-1})$ is obtained from $H$ by replacing 
each vertex $v_i$ of $H$ with $G_i$ for $i=0\ldots n-1$ and joining a vertex $x$ in $G_i$ to a vertex $y$ of $G_j$ if and only if $v_i$ and $v_j$ are adjacent in $H$. 
The graph $G_i$, $i=0\ldots n-1$ is said to be the {\em component} of the expansion associated to $v_i$.

\ligne In \cite{AraKarSub11}, Aravind et al proved that Conjecture \ref{Conj:Reed} holds true for {\em full} expansions and {\em independent} expansions of odd holes, that is expansions $H(G_1\ldots G_n)$ 
of odd holes where all the $G_i$'s are either complete graphs or edgeless graphs. Moreover, they ask for proving Conjecture \ref{Conj:Reed} for graph expansions whenever every component
 of the expansion statifies Conjecture \ref{Conj:Reed}.   

In this paper, we consider Conjecture \ref{Conj:Reed} for expansion of bipartite graphs, namely bipartite expansions and odd hole expansions. We use for this a colouring  algorithm of bipartite expansions 
that we extend to odd hole expansions, this allows us to compute the chromatic number of those graphs. We prove that Conjecture \ref{Conj:Reed} holds for a bipartite expansion (Theorem \ref{Theorem:ReedsForBipartiteExpansion}). 

\indent Moreover, Conjecture \ref{Conj:Reed} holds for odd hole expansions when the minimum chromatic number of the components is even (Corollary \ref{Corollary:ReedQuandMinChromaticEstPair}),
when some component of the expansion has chromatic number $1$ (Theorem 18),
or when a component induces a bipartite graph (Theorem \ref{Thm:OddExpansionComposanteBipartite}).
It is also the case  if all components have the same chromatic number (Theorem \ref{Thm:OddExpansionComposanteEqualChromaticNumber}),
if the components have chromatic number at most $4$ (Theorem \ref{Thm:OddExpansionComposanteAtMost4}), and when the odd hole has length $5$ (Theorem \ref{Theorem:C5_Expansion}).
In addition, if $G$ is an odd hole expansion we have $\chi(G)\leq\lceil\frac{\omega(G)+\Delta(G)+1}{2}\rceil+1$ (Theorem \ref{Thm:Reed_à_2_Près}).

These results improve the result of Aravind et al on full and independent expansions of odd holes. 

The present section ends with some notations and preliminary results. Section \ref{Sec:ColouringBipartiteExpansions} is devoted to the colouring of bipartite expansions and its consequences on 
Conjecture \ref{Conj:Reed} for such graphs while in Section \ref{Sec:OddHoleExpansionsColouring} we consider the colouring of odd hole expansions and its implications on Conjecture \ref{Conj:Reed} 
are considered in Section \ref{sec:Applications}.

\subsection{Notations and preliminary results}
Given a graph $G$ and $X$ a subset of its vertex set, we denote $G[X]$ the subgraph of $G$ induced by $X$. The degree of a vertex $v$ in the graph $G$ is denoted $d_G(v)$ or $d(v)$ when no confusion is possible.
For an expansion  $H(G_0\ldots G_{n-1})$ of some graph $H$, we will assume in the following that the vertices of $H$ are weighted with the chromatic number of their associated component 
while an edge of $H$ is weighted with the sum of the weights of its endpoints.
Moreover, for $i=0,\ldots {n-1}$, we will denote $\chi_i$ as the chromatic number of $G_i$, while $V_i$ is for the vertex set of $G_i$, $\Delta_{i}$ is the maximum degree of $G_i$, and $\omega_i$ its clique number.

\begin{Lem}\label{Lemma:ReedPourSousGrapheQuiAtteintLeChromaticNumber}
 Let $H$ be an induced subgraph of some graph $G$ such that \\ $\chi(H)=\chi(G)$.
If $\chi(H)\leq\lceil\frac{\omega(H)+\Delta(H)+1}{2}\rceil$ then $\chi(G)\leq\lceil\frac{\omega(G)+\Delta(G)+1}{2}\rceil$.
\end{Lem}
\begin{Prf}
 Since $H$ is an induced  subgraph of $G$, $\omega(H)\leq \omega(G)$ and $\Delta(H)\leq\Delta(G)$.
Thus $\chi(G)=\chi(H)\leq\lceil\frac{\omega(H)+\Delta(H)+1}{2}\rceil\leq\lceil\frac{\omega(G)+\Delta(G)+1}{2}\rceil$.
\end{Prf}

\begin{Thm}\cite{Rab08} \label{Theorem:Complement}If $\overline{G}$ is disconnected then
$\chi(G) \leq \lceil \frac{\omega(G) + \Delta(G)+1}{2}\rceil$
\end{Thm}

\begin{Lem}\label{Lemma:Connected_component_odd_hole_expansion_minimum_cexemple}
Let $G=H(G_0\ldots G_{n-1})$ be an odd hole expansion that is a minimum counter-example of Conjecture \ref{Conj:Reed} (if any). For $i\in\{0\ldots n\}$, $G_{i}$ is connected.
\end{Lem}
\begin{Prf} Without loss of generality assume that the subgraph induced by $G_0$ is not connected. Let $X_{1}$ and $X_{2}$ be two subset of $V(G_0)$ inducing a connected component and suppose that we need at
 most $\chi_{j}$ colors ($j=1,2$) to color $X_{j}$ with $\chi_{1} \leq \chi_{2}$. Let $G^{'}$ be the subgraph obtained from  $G$ by deleting $X_{1}$. 
Since $G^{'}$ satisfies Conjecture \ref{Conj:Reed} by hypothesis, we have 
$\chi(G^{'}) \leq \lceil \frac{\Delta(G^{'}) + \omega(G^{'})+1}{2}\rceil$. We can then color the vertices of $X_{1}$ by using the colors appearing in $X_{2}$ since $\chi_{1} \leq \chi_{2}$. 
Since $\omega(G) \geq \omega(G^{'})$ and $\Delta(G) \geq \Delta(G^{'})$, we have \\
$\chi(G) = \chi(G^{'}) \leq \lceil \frac{\omega(G^{'}) + \Delta(G^{'})+1}{2}\rceil \leq  \lceil \frac{\omega(G) + \Delta(G)+1}{2}\rceil$, a contradiction.
\end{Prf}

\section{Coloring of bipartite expansion}\label{Sec:ColouringBipartiteExpansions}
\begin{Nots}\label{Not:BipartiteExpansion}
Let $H$ be a bipartite graph with $n$ vertices: $v_{0},\ldots, v_{n-1}$ and $H(G_0\ldots G_{n-1})$ be an expansion of $H$.
Without loss of generality we assume that $v_{0}$ and $v_{1}$ are adjacent and are such that the edge $v_0v_1$ has maximum weight in $H$.\\
Let $\Gamma_{0}$ be a set of $\chi_{0}$ colors and $\Gamma_{1}$ be a set of $\chi_{1}$ other colors.\\
A given index $i\in\{0,\ldots n-1\}$ will have a {\em prefered} index in $\{0,1\}$, say $p(i)$, defined as follows~: $p(i)=0$ whenever $v_i$ and $v_0$ are vertices of the same class of the bipartition 
otherwise $p(i)$ will be defined to be $1$. Moreover we define the index $p'(i)$ such that $\{p(i),p'(i)\}=\{0,1\}$.
\end{Nots}

When $H(G_0\ldots G_{n-1})$ is a bipartite expansion, according to the above notations, $G_i$ ($0\leq i\leq n-1$) will be colored by using preferably the set of colors $\Gamma_{p(i)}$. 

More precisely $G_i$ will be colored by using  $Min(\chi_i,\chi_{p(i)})$ colors of $\Gamma_{p(i)}$ and $Max(0,\chi_i-\chi_{p(i)})$ colors of $\Gamma_{p'(i)}$ (see Theorem \ref{Thm:BipartiteExpansionColoring}).
\begin{Thm}\label{Thm:BipartiteExpansionColoring}
 Let $H(G_0\ldots G_{n-1})$ be a bipartite expansion. \\
For $i\in\{0\ldots n-1\}$,
\begin{description}
 \item if $\chi_i\leq \chi_{p(i)}$  then $G_i$ can be colored by using $\chi_i$ colors of $\Gamma_{p(i)}$
 \item otherwise $G_i$ can be colored by using the $\chi_{p(i)}$ colors of $\Gamma_{p(i)}$ together with $\chi_i-\chi_{p(i)}$ colors of $\Gamma_{p'(i)}$.
\end{description}
\end{Thm}
\begin{Prf}
 Let us colour the vertices of $G_{0}$ with the $\chi_{0}$ colors of $\Gamma_0$.
In the same way we colour the vertices of $G_{1}$ by using
the $\chi_{1}$ colors of $\Gamma_1$ (recall that $\Gamma_0\cap \Gamma_1=\emptyset$).

For ${i}\in \{2\ldots n-1\}$, we color the graph $G_i$ as follows~: when $\chi_{i} \leq \chi_{p(i)}$ we can use the $\chi_{i}$ first
colors in $\Gamma_{p(i)}$ to colour $G_{i}$~; and when $\chi_{i}> \chi_{p(i)}$ we color the vertices of $G_{i}$ by using
the $\chi_{p(i)}$ colors of $\Gamma_{p(i)}$ and the $\chi_{i}-\chi_{p(i)}$ last colors of $\Gamma_{p'(i)}$.

We claim that the resulting coloring is a proper coloring of $H(G_{0}, \ldots , G_{n-1})$.

Indeed let $v_{i}v_{j}$ be an edge of $H$. Let us remark first that we do not have $\chi_{i} > \chi_{p(i)}$ and $ \chi_{j} > \chi_{p(j)}$ since $\chi_{i} + \chi_{j} \leq
\chi_{0} + \chi_{1}$ by hypothesis, moreover, since $v_i$ and $v_j$ are adjacent we have $p(i)=p'(j)$ and $p(j)=p'(i)$.

{\bf case 1} {\em $\chi_{i} \leq \chi_{p(i)}$ and $ \chi_{j} \leq
 \chi_{p(j)}$}

The colors used in $G_{i}$  are only colors of $\Gamma_{p(i)}$ and those of
$G_{j}$ are only colors of $\Gamma_{p(j)}=\Gamma_{p'(i)}$ and these two sets of colours are disjoint.

{\bf case 2} {\em $\chi_{i} \leq \chi_{p(i)}$ and $ \chi_{j} > \chi_{p(j)}$}

The colours used in the coloring of $G_{i}$ are only the $\chi_i$ first colors of $\Gamma_{p(i)}$. In order to color $G_j$, we use all the
$\chi_{p(j)}$ colours of $\Gamma_{p(j)}$  and we need to use the last $\chi_{j}-\chi_{p(j)}$ colors of $\Gamma_{p'(j)}$.
Since $\chi_{i} + \chi_{j} \leq \chi_{0} + \chi_{1}=\chi_{p(j)}+\chi_{p(i)}$, we have $\chi_{j}-\chi_{p(j)} \leq \chi_{p(i)} - \chi_{i}$. Hence the set of
colors of $\Gamma_{p(i)}$ used in order to achieve the colouring of $G_{j}$ is disjoint from the set of colors used in $G_{i}$.

{\bf case 3} {\em $\chi_{i} > \chi_{p(i)}$ and $ \chi_{j} \leq
\chi_{p(j)}$} The same argument works.
\end{Prf}

From Theorem \ref{Thm:BipartiteExpansionColoring} and according to Notations \ref{Not:BipartiteExpansion}, since $\chi(H)\geq \chi_0+\chi_1$, we have:
\begin{Cor}\label{Cor:BipartiteExpansionChromaticNumber}
 Let $G=H(G_0\ldots G_{n-1})$ be a bipartite expansion, $\chi(G))=\chi_0+\chi_1$.
\end{Cor}
\begin{Rem}\label{Rem:PasDeDepassementDuNombreChromatique}
 Let us remark that the coloring given in Theorem \ref{Thm:BipartiteExpansionColoring} has the following property: $|\Gamma_i|=\chi_i$ for $i\in \{0\ldots n-1\}$. 
\end{Rem}

\begin{Thm} \label{Theorem:ReedsForBipartiteExpansion}
Any expansion of a bipartite graph satisfies Conjecture \ref{Conj:Reed}.
\end{Thm}
\begin{Prf}
Let $H(G_0\ldots G_{n-1})$ be an expansion of a bipartite graph $H$. According to Notations \ref{Not:BipartiteExpansion} and by Theorem \ref{Theorem:Complement}, the subgraph induced 
by $V(G_0) \cup V(G_1)$, say $G'$, verifies Conjecture \ref{Conj:Reed}. Moreover 
$\chi(G')=\chi_0+\chi_1$ and by Corollary \ref{Cor:BipartiteExpansionChromaticNumber} $\chi(G)=\chi(G')$. The result follows from Lemma \ref{Lemma:ReedPourSousGrapheQuiAtteintLeChromaticNumber}.
\end{Prf}

\section{Odd hole expansions coloring}\label{Sec:OddHoleExpansionsColouring}
By Theorem \ref{Theorem:Complement}, an expansion of triangle verifies Conjecture \ref{Conj:Reed}.
In what follows $C_{2k+1}$ denotes an odd hole of length $2k+1$ ($k\geq 2$) and all indexes are taken modulo $2k+1$.
Moreover, the vertex set of $C_{2k+1}$ is $\{v_0,\ldots v_{2k}\}$ and $v_iv_j$ is an edge if and only if $j=i+1$.

Theorem \ref{Theorem:ColoriageExpansionOddHole} below provides a proper coloring for odd hole expansions.
\begin{Thm}\label{Theorem:ColoriageExpansionOddHole}
Let $G=C_{2k+1}(G_0\ldots G_{2k})$ be an expansion of an odd hole. 
Assume that the edge $v_0v_1$ has maximum weight in $H$.\\
Let $i$ be an index in $\{3\ldots 2k-1\}$.
\begin{description}
 \item If $\chi_{0}+\chi_{1}\geq \chi_{i-1}+\chi_{i}+\chi_{i+1}$ then $\chi(G)\leq\chi_{0}+\chi_{1}$
 \item \begin{description}
	\item else if $\chi_{i-1}>\chi_{p(i-1)}$ and $\chi_{i+1}>\chi_{p(i+1)}$ then  $\chi(G)\leq\chi_{0}+\chi_{1}+\lfloor\frac{\chi_i+1}{2}\rfloor$
	\item else $\chi(G)\leq\chi_{0}+\chi_{1}+\lfloor\frac{\chi_{i-1}+\chi_i+\chi_{i+1}-\chi_{0}-\chi_{1}+1}{2}\rfloor$.
       \end{description}
\end{description}

\end{Thm}
\begin{Prf}

Let $H'$ be the bipartite graph whose vertex set is $V(C_{2k+1})-\{v_i\}$. Assume that the coloring described in Theorem \ref{Thm:BipartiteExpansionColoring} has been 
applied to the expansion $H'(G_0,G_1\ldots G_{i-1}, G_{i+1}\ldots G_{2k})$. Observe that the notations $p(i)$ and $p'(i)$ are not defined in $C_{2k+1}(G_0\ldots G_{2k})$, 
however in the following we will use 
this notations as meant in $H'(G_0,G_1\ldots G_{i-1}, G_{i+1}\ldots G_{2k})$, thus we have $p(i-1)=p'(i+1)$ 
and $p'(i-1)=p(i+1)$.

\ligne\noindent Let us now consider the coloring of $G_i$. 

According to the coloring of $G_{i-1}$ and those of $G_{i+1}$ four cases may occur.

{\em Case $1$~: $\chi_{i-1}\leq \chi_{p(i-1)}$ and  $\chi_{i+1}\leq \chi_{p(i+1)}$.} The coloring of $G_{i-1}$ uses $\chi_{i-1}$ colors of ${\Gamma}_{p(i-1)}$ and none 
in the set ${\Gamma}_{p'(i-1)}$ while the coloring of $G_{i+1}$ needs only $\chi_{i+1}$ colors in ${\Gamma}_{p(i+1)}$~; consequently there are $\chi_0+\chi_1-\chi_{i-1}-\chi_{i+1}$ colors free in 
${\Gamma}_0\cup {\Gamma}_1$ for the coloring of $G_i$.

{\em Case $2$~: $\chi_{i-1}\leq \chi_{p(i-1)}$ and  $\chi_{i+1}> \chi_{p(i+1)}$.} We have the same coloring for $G_{i-1}$ as in {\em Case $1$}. But the subgraph $G_{i+1}$ is colored with all 
the colors of ${\Gamma}_{p(i+1)}$ together with $\chi_{i+1}-\chi_{p(i+1)}$ colors of ${\Gamma}_{p'(i+1)}$, once again there are in ${\Gamma}_{p'(i+1)}$ at least 
$\chi_{p'(i+1)}-\chi_{i-1}-(\chi_{i+1}-\chi_{p(i+1)})$ free colors for the coloring of $G_i$. 

{\em Case $3$~:$\chi_{i-1}> \chi_{p(i-1)}$ and  $\chi_{i+1}\leq \chi_{p(i+1)}$.} We color $G_{i-1}$ with the $\chi_{p(i-1)}$ colors of ${\Gamma}_{p(i-1)}$ and 
with $\chi_{i-1}-\chi_{p(i-1)}$ colors of ${\Gamma}_{p'(i-1)}$. The subgraph $G_{i+1}$ being colored with $\chi_{i+1}$ colors in ${\Gamma}_{p(i+1)}$. 
Thus there are $\chi_{p'(i-1)}-\chi_{i+1}-(\chi_{i-1}-\chi_{p(i-1)})$ unused colors in ${\Gamma}_{p'(i-1)}$.

{\em Case $4$~: $\chi_{i-1}> \chi_{p(i-1)}$ and  $\chi_{i+1}> \chi_{p(i+1)}$.} In this case $G_{i-1}]$ can be colored with all the colors in ${\Gamma}_{p(i-1)}$ and $\chi_{i-1}-\chi_{p(i-1)}$ colors
of ${\Gamma}_{p'(i-1)}$. Moreover the coloring of $G_{i+1}$ is done with the colors of ${\Gamma}_{p(i+1)}$ and $\chi_{i+1}-\chi_{p(i+1)}$ additionnal colors of ${\Gamma}_{p'(i+1)}$. All colors of 
${\Gamma}_{0}\cup {\Gamma}_{1}$ are used in this colorings, but just observe that $\chi_i<Min(\chi_{0},\chi_{1})$.

\ligne \noindent Suppose first $\chi_{0}+\chi_{1}\geq \chi_{i-1}+\chi_{i}+\chi_{i+1}$. 

In this situation {\em Case $4$} cannot occur and  there are enough free colors in ${\Gamma}_{0}\cup {\Gamma}_{1}$ for the coloring of $G_i$. Hence $\chi(G)\leq\chi_{0}+\chi_{1}$.

\ligne \noindent  From now on $\chi_{0}+\chi_{1}<\chi_{i-1}+\chi_{i}+\chi_{i+1}$. 

\ligne Assume now  $\chi_{i-1}> \chi_{p(i-1)}$ and  $\chi_{i+1}> \chi_{p(i+1)}$. Recall that $\chi_i<\chi_{0}$ and $\chi_i<\chi_{1}$. 
Let $a=\lfloor\frac{\chi_i}{2}\rfloor$ and ${\Gamma}$ be a set of $a$ additionnal colors. The coloring of $G{i-1}$ uses $\chi_{p(i-1)}$ colors of ${\Gamma}_{p(i-1)}$, let use replace $a$ of those colors with 
the colors of ${\Gamma}$. 
We also replace $a$ colors  of ${\Gamma}_{p(i+1)}$ with the same colors of ${\Gamma}$. Thus $2a$ colors are left for the coloring of $G_i$, that is $\chi_i$ or $\chi_i-1$ according to the parity of $\chi_i$.
Hence in this case the whole graph can be colored with at most $|{\Gamma}_{0}|+|{\Gamma}_{1}|+a+1$ colors, that is $\chi(G)\leq \chi_{0}+\chi_{1}+\lfloor\frac{\chi_i+1}{2}\rfloor$.

\ligne Finally, assume $\chi_{i-1}\leq \chi_{p(i-1)}$ or  $\chi_{i+1}\leq \chi_{p(i+1)}$. Recall that there are $\chi_{0}+\chi_{1}-\chi_{i-1}-\chi_{i+1}$ colors free in 
${\Gamma}_{0}\cup {\Gamma}_{1}$  for the coloring of $G_i$.
Since $\chi_{0}+\chi_{1}\geq \chi_i+\chi_{i-1}$ it is clear that $\chi_{i+1}\geq \frac{\chi_{i-1}+\chi_i+\chi{i+1}-\chi_{0}-\chi_{1}}{2}$. 
Similarly $\chi_{i-1}\geq \frac{\chi_{i-1}+\chi_i+\chi{i+1}-\chi_{0}-\chi_{1}}{2}$. Let us state $a=\lfloor\frac{\chi_{i-1}+\chi_{i}+\chi_{i+1}-\chi_{0}-\chi_{1}}{2}\rfloor$  
and ${\Gamma}$ be a set of $a$ additionnal colors.
We replace, in the coloring of $G_{i-1}$, \Ajoute{a number of }$a$ colors of $\Gamma_{p(i-1)}$ with the colors of ${\Gamma}$ as well as $a$ colors of ${\Gamma}_{p(i+1)}$ in the coloring of $G_{i+1}$. 
Hence we have $2a$ more colors for the coloring of $G_i$. 
It follows that the whole graph can be colored with the colors of ${\Gamma}_{0}\cup {\Gamma}_{1}\cup {\Gamma}$ and possibly an additionnal color according to the parity 
of $\chi_{i+1}+\chi_{i}+\chi_{i-1}-\chi_{0}-\chi_{1}$.
Thus in this case $\chi(G) \leq\chi_{0}+\chi_{1}+\lfloor\frac{\chi_{i-1}+\chi_i+\chi_{l+1}-\chi_{0}-\chi_{1}+1}{2}\rfloor$.
\end{Prf}

Theorem \ref{Theorem:ChromaticNumberExpansionOddHole} gives the chromatic number for odd hole expansions.
\begin{Thm}\label{Theorem:ChromaticNumberExpansionOddHole}
Let $G=C_{2k+1}(G_0\ldots G_{2k})$ be an expansion of an odd hole. 
We assume that the edge $v_0v_1$ has maximum weight in $C_{2k+1}$.\\
Let $l$ be an index such that
$\chi_{l-1}+\chi_l+\chi_{l+1}=\begin{matrix}\; \\ Min \\  _{3\leq i\leq 2k-1}\end{matrix} \left \{\chi_{i-1}+\chi_i+\chi_{i+1}\right \}$
\begin{description}
 \item If $\chi_{0}+\chi_{1}\geq \chi_{l-1}+\chi_{l}+\chi_{l+1}$ then $\chi(G)=\chi_{0}+\chi_{1}$
 \item else $\chi(G)=\chi_{0}+\chi_{1}+\lfloor\frac{\chi_{l-1}+\chi_l+\chi_{l+1}-\chi_{0}-\chi_{1}+1}{2}\rfloor$.
\end{description}
\end{Thm}
\begin{Prf}
 Since $\chi(G)\geq \chi_0+\chi_1$, by Theorem \ref{Theorem:ColoriageExpansionOddHole} we can suppose that $\chi_{0}+\chi_{1}< \chi_{l-1}+\chi_{l}+\chi_{l+1}$.

In addition  $\chi_{l-1}\leq\chi_{p(l-1)}$ or  $\chi_{l+1}> \chi_{p(l+1)}$. Otherwise, since $\chi_{l-1}>\chi_{p(l-1)}$ we have $l-1>2$ and then $\chi_{l-2}<\chi_{p(l-2)}=\chi_{p(l+1)}<\chi_{l+1}$. 
It follows that $\chi_l+\chi_{l-1}+\chi_{l-2}<\chi_{l+1}+\chi_l+\chi_{l-1}$, a contradiction with the choice of the index $l$. 

Hence by Theorem \ref{Theorem:ColoriageExpansionOddHole} we have 
$\chi(G)\leq\chi_{0}+\chi_{1}+\lfloor\frac{\chi_{l-1}+\chi_l+\chi_{l+1}-\chi_{0}-\chi_{1}+1}{2}\rfloor$ and there is a coloring of $G$ using colors in $\Gamma_0\cup \Gamma_1\cup \Gamma$ where $\Gamma_0$, $\Gamma_1$ and $\Gamma$ are disjoint sets of colors such 
that $|\Gamma_{0}|=\chi_0$, $|\Gamma_{1}|=\chi_1$ and $|\Gamma|=\lfloor\frac{\chi_{l-1}+\chi_l+\chi_{l+1}-\chi_{0}-\chi_{1}+1}{2}\rfloor$. Since the sum $\chi_{l+1}+\chi_l+\chi_{l-1}$ is minimum, Theorem \ref{Theorem:ColoriageExpansionOddHole}
cannot provide a coloring using less colors.

\ligne Assume now $\chi(G)<|\Gamma_{0}|+|\Gamma_{1}|+|\Gamma|$. We can suppose that an optimal coloring of $G$ uses the set $\Gamma_{0}\cup \Gamma_{1}\cup \Gamma'$ as set of colors 
where $\Gamma'\cap(\Gamma_{0}\cup \Gamma_{1})=\emptyset$   and $|\Gamma'|<|\Gamma|$. In a such coloring the number of unused colors for the coloring of $X_{l+1}$ and $X_{l-1}$ is 
at most $\chi_{0}+\chi_{1}+|\Gamma'|-\chi_{l+1}-\chi_{l-1}$.

Thus $\chi_l\leq \chi_{0}+\chi_{1}+|\Gamma'|-\chi_{l+1}-\chi_{l-1}$  and then 
$$\chi_l+\chi_{l+1}+\chi_{l-1}-\chi_{0}-\chi_{1}\leq|\Gamma'|<\lfloor\frac{\chi_l+\chi_{l+1}+\chi_{l-1}-\chi_{0}-\chi_{1}+1}{2}\rfloor,$$ a contradiction with the fact 
that $\chi_l+\chi_{l+1}+\chi_{l-1}-\chi_{0}-\chi_{1}$ is a positive integer.
\end{Prf}
\section{Applications}\label{sec:Applications}
In \cite{AraKarSub11} Aravind et al observed that the complete or independent expansions of an odd hole satisfy  Conjecture \ref{Conj:Reed}. We give below improvements of this results.

\begin{Cor}\label{Cor:ReedsConjecturePourOddHole}
 Conjecture \ref{Conj:Reed} holds for an odd hole expansion when, in the conditions of Theorem \ref{Theorem:ColoriageExpansionOddHole}, we have $\chi(A)=\omega(A)$ for $A\in \{G_{0},G_{1}, G_l\}$.
\end{Cor}
\begin{Prf}
By Theorem \ref{Theorem:ChromaticNumberExpansionOddHole} we know that $\chi(G)\leq\frac{\chi_0+\chi_1+\chi_{l-1}+\chi_{l}+\chi_{l+1} +1}{2}$.
By assumption we have $\chi_0+\chi_1=\omega(G_{0})+\omega(G_{1})\leq\omega(G)$, moreover if $v$ is a vertex of a maximum clique in $G_l$, 
$d(v)\geq \omega(G_l)-1+|V_{l+1}|+|V_{l-1}|$ then $\Delta\geq \chi_l+\chi_{l+1}+\chi_{l-1}-1$. The result follows.
\end{Prf}
\begin{Cor}\label{Corollary:UpperBoundChromaticNumberExpansionOddHole}
Let $G=C_{2k+1}(G_0\ldots G_{2k})$ be an expansion of an odd hole. 
Let $\displaystyle{p=\min_{0 \leq i \leq 2k} \chi_{i}}$. Assume that the edge $v_iv_{i+1}$ has maximum weight in $C_{2k+1}$ for some $i\in\{0,\ldots 2k\}$. 
Then $\chi(G) \leq \chi_{i}+\chi_{i+1}+\lfloor\frac{p+1}{2}\rfloor$.
\end{Cor}

\begin{Prf}

By Theorem \ref{Theorem:ColoriageExpansionOddHole}, we may assume for $j\in\{i+3,i+4\ldots i-2\}$
\begin{equation}\label{Eq:Cas3Thm9}
  \chi(G)\leq\chi_{i}+\chi_{i+1}+\lfloor\frac{\chi_{j-1}+\chi_j+\chi_{j+1}-\chi_{i}-\chi_{i+1}+1}{2}\rfloor.
\end{equation}
Moreover, there is an index $l\in\{i+2,\ldots i-1\}$ such that $\chi_l=p$, otherwise $\chi_i=p$ or $\chi_{i+1}=p$. Suppose without loss of generality $\chi_{i+1}=p$. 
But now, since $\chi_{i-1}>\chi_{i+1}$ 
we have $\chi_{i-1}+\chi_i>\chi_i+\chi_{i+1}$, a contradiction, since the edge $v_iv_{i+1}$ has maximum weight in $C_{2k+1}$.

If $l\geq i+4$, we apply (\ref{Eq:Cas3Thm9}) with $j=l-1$, since $\chi_{l-1}+\chi_{l-2}\leq \chi_i+\chi_{i+1}$ we get 
$\chi(G)\leq\chi_{i}+\chi_{i+1}+\lfloor\frac{\chi_{l-2}+\chi_{l-1}+\chi_{l}-\chi_{i}-\chi_{i+1}+1}{2}\rfloor\leq \chi_{i}+\chi_{i+1}+\lfloor\frac{\chi_{l}+1}{2}\rfloor.$

If $l=i+2$ or $l=i+3$, we apply (\ref{Eq:Cas3Thm9}) with $j=l+1$, since  $\chi_{l+1}+\chi_{l+2}\leq \chi_i+\chi_{i+1}$ we get
$\chi(G)\leq\chi_{i}+\chi_{i+1}+\lfloor\frac{\chi_{l+2}+\chi_{l+1}+\chi_{l}-\chi_{i}-\chi_{i+1}+1}{2}\rfloor\leq \chi_{i}+\chi_{i+1}+\lfloor\frac{\chi_{l}+1}{2}\rfloor.$
 In both cases, it follows $\chi(G)\leq \chi_{i}+\chi_{i+1}+\lfloor\frac{p+1}{2}\rfloor$.
\end{Prf}

\begin{Cor}\label{Cor:OtherNeighborhoodEdgeMaximumWeight}
Let $G=C_{2k+1}(G_0\ldots G_{2k})$ be an expansion of an odd hole. Let $v_iv_{i+1}$ be an edge of maximal weight in $C_{2k+1}$. 
Assume that $\chi(G)=\chi_{i}+\chi_{i+1}+q+1$ for some integer $q\geq 0$.
If $G[V_i\cup V_{i+1}]$ has a vertex of maximum degree in $V_i$ (resp. $V_{i+1}$) then either Conjecture \ref{Conj:Reed} holds for $G$ or 
$V_{i-1}$ (resp.$V_{i+2}$) induces a graph on at most $2q+1$ vertices.
\end{Cor}

\begin{Prf}
Assume that $G$ is a counter-example to Conjecture \ref{Conj:Reed}. For convenience we note $G'=G[V_i\cup V_{i+1}]$.

Let $v$ be a vertex of maximum degree in $G'$, suppose that $v\in V_i$ and $V_{i-1}$ has at least $2q+2$ vertices.

We have $\Delta(G)\geq d_{G'}(v)+|V_{i-1}| \geq\Delta(G')+2q+2$ and $\omega(G)\geq \omega(G')$.
Thus $\lceil\frac{\omega(G)+\Delta(G)+1}{2}\rceil\geq \lceil\frac{\omega(G')+\Delta(G')+1+2q+2}{2}\rceil=\lceil\frac{\omega(G')+\Delta(G')+1}{2}\rceil+q+1$.

Since by Theorem  \ref{Theorem:Complement}, $G'$ verifies Conjecture \ref{Conj:Reed}, $\lceil\frac{\omega(G')+\Delta(G')+1}{2}\rceil\geq \chi_i+\chi_{i+1}$. 
Hence by Corollary \ref{Corollary:UpperBoundChromaticNumberExpansionOddHole}, 
$\lceil\frac{\omega(G)+\Delta(G)+1}{2}\rceil\geq \chi_i+\chi_{i+1}+q+1 = \chi(G)$, a contradiction. 
\end{Prf}
\begin{Cor}\label{Corollary:ChromaticNumberExpansionOddHole_If_Couterexample}
Let $G=C_{2k+1}(G_0\ldots G_{2k})$ be an expansion of an odd hole and let $\displaystyle{p=\min_{0
\leq i \leq 2k} \chi_{i}}$. If  the edge $v_{i}v_{i+1}$ has maximum weight
in $C_{2k+1}$ then Conjecture \ref{Conj:Reed} holds for $G$ or $\chi(G) =
\chi_{i}+\chi_{i+1}+\lfloor\frac{p+1}{2}\rfloor$.
\end{Cor}
\begin{Prf}
We know by Corollary \ref{Corollary:UpperBoundChromaticNumberExpansionOddHole} that $\chi(G) \leq\chi_{i}+\chi_{i+1}+\lfloor\frac{p+1}{2}\rfloor$. Assume that $G$ is a counter-example to Conjecture \ref{Conj:Reed}
 and $\chi(G) \neq\chi_{i}+\chi_{i+1}+\lfloor\frac{p+1}{2}\rfloor$. 
Thus, we have $\chi(G) \leq\chi_{i}+\chi_{i+1}+\lfloor\frac{p}{2}\rfloor$.\\
Assume without loss of generality  that $v \in V_{i+1}$ is a vertex with maximum degree in $G^{'}=G[V_{i} \cup V_{i+1}]$. By Theorem \ref{Theorem:Complement}, $G^{'}$ satisfies Conjecture \ref{Conj:Reed}.

Hence $\lceil\frac{\omega(G^{'})+\Delta(G^{'})+1}{2}\rceil\geq \chi_{i}+\chi_{i+1}=\chi(G^{'})$.
Since $G_{i+2}$ has at least $p$ vertices, we have $\Delta(G) \geq d(v) \geq |V_{i}|+\Delta_{i+1}+p \geq \Delta(G^{'})+p$, which leads to  

$\lceil\frac{\omega(G)+\Delta(G)+1}{2}\rceil\geq \lceil\frac{\omega(G^{'})+\Delta(G^{'})+p+1}{2}\rceil\geq \lceil\frac{\omega(G^{'})+\Delta(G^{'})+1}{2}\rceil+\lfloor\frac{p}{2}\rfloor$.

Hence $\lceil\frac{\omega(G)+\Delta(G)+1}{2}\rceil\geq \chi(G)$, a contradiction.
\end{Prf}

\begin{Thm}\label{Theorem:NeighborhoodEdgeMaximumWeight}
Let $G=C_{2k+1}(G_0\ldots G_{2k})$ be an expansion of an odd hole of length $2k+1$ and let $\displaystyle{p=\min_{0\leq i \leq 2k} \chi_{i}}$. Let $v_iv_{i+1}$ be an edge of maximal
weight in $C_{2k+1}$ and assume that $v \in V_{i+1}$ is a vertex of
maximum degree in $G^{'}=G[V_i\cup V_{i+1}]$.
If $G$ does not satisfy Conjecture \ref{Conj:Reed} then
$V_{i+2}$ induces a complete graph on $p$ vertices and
$v_{i+3}v_{i+4}$ is an edge of maximal weight in $C_{2k+1}$.
\end{Thm}

\begin{Prf}
Assume that $G$ does not satisfy Conjecture \ref{Conj:Reed}
and $V_{i+2}$ does not induce a complete graph on $p$ vertices. By Corollary
\ref{Corollary:ChromaticNumberExpansionOddHole_If_Couterexample}, we have $\chi(G) =\chi_{i}+\chi_{i+1}+\lfloor\frac{p+1}{2}\rfloor$.

We may assume that $|V_{i+2}|\geq p+1$ otherwise $V_{i+2}$ would induce a complete graph on $p$ vertices, a contradiction.

We have $\Delta(G)\geq d_{G'}(v)+|V_{i+2}| \geq\Delta(G^{'})+p+1$ and $\omega(G)\geq \omega(G^{'})$.

Hence
$\lceil\frac{\omega(G)+\Delta(G)+1}{2}\rceil\geq\lceil\frac{\omega(G^{'})+\Delta(G^{'})+p+2}{2}\rceil\geq\lceil\frac{\omega(G^{'})+\Delta(G^{'})+1}{2}\rceil+\lfloor\frac{p+1}{2}\rfloor=\chi(G)$, a contradiction.

Assume now that $\chi_{i+3}+\chi_{i+4} \leq \chi_{i}+\chi_{i+1}-1$. By Theorem \ref{Theorem:ChromaticNumberExpansionOddHole} we have
$\chi(G)\leq\chi_{i}+\chi_{i+1}+\lfloor\frac{\chi_{i+2}+\chi_{i+3}+\chi_{4}-\chi_{i}-\chi_{i+1}+1}{2}\rfloor$
which leads to  $\chi(G)\leq\chi_{i}+\chi_{i+1}+\lfloor\frac{p}{2}\rfloor$. Moreover, $\Delta(G)\geq d_{G'}(v)+|V_{i+2}| \geq\Delta(G^{'})+p$ and $\omega(G)\geq \omega(G^{'})$. 
Hence, $\lceil\frac{\omega(G)+\Delta(G)+1}{2}\rceil\geq\lceil\frac{\omega(G^{'})+\Delta(G^{'})+p+1}{2}\rceil\geq\lceil\frac{\omega(G^{'})+\Delta(G^{'})+1}{2}\rceil+\lfloor\frac{p}{2}\rfloor\geq\chi(G)$, a contradiction. 
Henceforth $v_{i+3}v_{i+4}$ is an edge of maximum weight in $C_{2k+1}$ as claimed.
\end{Prf}
\begin{Cor}\label{Corollary:ReedQuandMinChromaticEstPair}
Let $G=C_{2k+1}(G_0\ldots G_{2k})$ be an expansion of an odd hole. Let $\displaystyle{p=\min_{0\leq i \leq 2k} \chi_{i}}$. 
If $p$ is even then Conjecture \ref{Conj:Reed} holds for $G$.
\end{Cor}

\begin{Prf}
Let us write $C_{2k+1}=v_0\ldots v_{2k}$. Suppose the edge $v_iv_{i+1}$ has maximum weight in $C_{2k+1}$. Let $G'=G[V_i\cup V_{i+1}]$ and $v$ be a vertex of maximum degree in $G'$. Assume without loss of genenality $v\in V_{i+1}$. 
Since $p$ is even,$\lfloor\frac{p+1}{2}\rfloor=\lfloor\frac{p}{2}\rfloor$ and  from Corollary \ref{Corollary:ChromaticNumberExpansionOddHole_If_Couterexample} we have:
$\chi(G) =\chi_{i}+\chi_{i+1}+\lfloor\frac{p}{2}\rfloor$. 
In addition, by Theorem \ref{Theorem:NeighborhoodEdgeMaximumWeight}, $V_{i+2}$ induces a complete graph on $p$ vertices. Thus, $\Delta(G)\geq d_{G'}(v)+|V_{i+2}| \geq\Delta(G^{'})+p$. 
Consequently,$\lceil\frac{\omega(G)+\Delta(G)+1}{2}\rceil\geq\lceil\frac{\omega(G^{'})+\Delta(G^{'})+p}{2}\rceil\geq\lceil\frac{\omega(G^{'})+\Delta(G^{'})+1}{2}\rceil+\lfloor\frac{p}{2}\rfloor=\chi(G)$,
 a contradiction.
\end{Prf}

\begin{Thm}\label{Thm:OddExpansionComposanteSingleton}
If $G=C_{2k+1}(G_0\ldots G_{2k})$ is an expansion of an odd hole  such that 
$\chi_i=1$ for some $i \in\{0\ldots 2k\}$ then Conjecture \ref{Conj:Reed} holds for $G$.
\end{Thm}
\begin{Prf}
Suppose that $G$ is a counter-example to Conjecture \ref{Conj:Reed}.
 
Assume, without loss of generality that $v_{0}v_{1}$ has maximum weight. 
By Corollary \ref{Corollary:UpperBoundChromaticNumberExpansionOddHole} we have $\chi(G) \leq \chi_{0}+\chi_{1}+1$. If $\chi(G) = \chi_{0}+\chi_{1}$ then $G$ satisfies 
Conjecture \ref{Conj:Reed} by Lemma \ref{Lemma:ReedPourSousGrapheQuiAtteintLeChromaticNumber}, a contradiction. Hence $\chi(G) = \chi_{0}+\chi_{1}+1$ and by 
Theorem \ref{Theorem:NeighborhoodEdgeMaximumWeight} we can suppose that $V_{2k}$ is reduced to a single vertex $v$.

We consider  an optimal coloring of the bipartite expansion $G-v$, such a coloring requires precisely $\chi_0+\chi_1$ colors and 
we can assume that this optimal coloring have been obtained via the algorithm described in the previous section (expansion of bipartite graphs). 
We denote $\Gamma_{i}$ the set of colors used for the coloring of
$G_i$, $i=0\ldots 2k-1$. When $i$ is even, $0$ is the preferred index for the coloring of $G_{i}$ and, $1$ is its preferred index when $i$ is odd. 
Let us remark that, for this coloring, when $i\in \{0\ldots 2k\}$, $\Gamma_{i}\cap \Gamma_{i+1}=\emptyset$, $\Gamma_{i} \subseteq \Gamma_{0} \cup \Gamma_{1}$, and $|\Gamma_i|=\chi_i$ 
(see Remark \ref{Rem:PasDeDepassementDuNombreChromatique}).
We get an optimal coloring of the whole graph $G$ by giving a new color to the vertex $v$.

\begin{Clm}\label{Clm:v2k-1v2k-2MaximumWeight}
$v_{2k-1}v_{2k-2}$ is an edge of maximum weight, moreover $\Gamma_{1} \subseteq \Gamma_{2k-1}$ and $\Gamma_{2k-2} \subseteq \Gamma_{0}$.
\end{Clm}
\begin{PrfClaim}
Suppose $\chi_{2k-1}<\chi_1$. Since $|\Gamma_{2k-1}|=\chi_{2k-1}$, some color $a$ of $\Gamma_{1}$ does not appear in $\Gamma_{2k-1}$. This color could be given to $v$, a contradiction.
Hence, $\chi_{2k-1}\geq\chi_1$, $\Gamma_{1} \subseteq \Gamma_{2k-1}$ and, consequently, $\Gamma_{2k-2} \subseteq \Gamma_{0}$.

If $\chi_{2k-2}<\chi_0$ then some color $a \in \Gamma_{0} \setminus \Gamma_{2k-2}$ does not appear in $\Gamma_{2k-1}$. 
Choose any color $b\in \Gamma_{1}$ and change the color of the vertices of $G_{2k-1}$, with that color, in $a$. Hence $b$ is now available to color $v$, a contradiction.

It follows $\chi_{2k-1}+\chi_{2k-2}\geq \chi_0+\chi_1$, that is the edge $v_{2k-1}v_{2k-2}$ has maximum weight.
\end{PrfClaim}

\begin{Clm}\label{Clm:GabConnected1}
Let $a$ be a color in $\Gamma_{2k-1} \cap \Gamma_{1}$ and $b$ be a color in $\Gamma_{2k-2}$. Then the subgraph $G_{ab}$ of $G$ induced by these two colors is connected.
\end{Clm}
\begin{PrfClaim} Let us remark that, by the definition of the expansion of an hole, it is sufficient to prove that $G_{ab}$ contains a vertex of color $b$ of $G_{0}$. 
Assume to the contrary that $G_{ab}$ is not connected. That is, the set of vertices colored with $b$ in $G_{0}$ is not contained in the connected component of $G_{ab}$ containing the vertices 
of color $a$ in $G_{2k-1}$. We can thus exchange the two colors $a$ and $b$ on the component containing the vertices of color $a$ in $G_{2k-1}$. 
Since $a$ does no longer appear in the neighborhood of $v$, we can give this color to $v$ and we get a $\chi_{0}+\chi_{1}$ coloring of $G$, a contradiction.

\end{PrfClaim}

\begin{Clm}\label{Clm:C1InC2i+1_C2iInC0}
For any $i$ ($0\leq i\leq 2k-1$), $\Gamma_{i} \subseteq \Gamma_{0}$ when $i$ is even and $\Gamma_{1} \subseteq \Gamma_{i}$ when $i$ is odd.
\end{Clm}
\begin{PrfClaim} Let $a$ be any color in $\Gamma_{2k-1} \cap \Gamma_{1}$ and $b$ any color in $\Gamma_{2k-2}$. Since by Claim \ref{Clm:GabConnected1}, $G_{ab}$ is connected, a shortest path in this subgraph joining a vertex in $G_{0}$ to a vertex in $G_{2k-1}$ must contain an edge between $G_{i}$ and $G_{i+1}$ for any index $i$ ($0\leq i\leq 2k-2$). Hence, when $i$ is even $G_{i}$ contains a vertex colored with $b$ ($0 \leq i \leq 2k-2$) while for $i$ odd $G_{i}$ contains a vertex colored with $a$ ($1 \leq i \leq 2k-1$). Since, by Claim \ref{Clm:v2k-1v2k-2MaximumWeight}, $\Gamma_{1} \subseteq \Gamma_{2k-1}$ and $\Gamma_{2k-2} \subseteq \Gamma_{0}$, the claim follows.
\end{PrfClaim}

\begin{Clm}\label{Clm:C2iDansC2i-2}
For any even index $i$ ( $2\leq i\leq 2k-2$), $\Gamma_{i} \subseteq \Gamma_{i-2}$.
\end{Clm}
\begin{PrfClaim}
Assume that some color $a$ of $\Gamma_{i}$ does not appear in $\Gamma_{i-2}$ and let $b$ be any color in $\Gamma_{1} \cap \Gamma_{2k-1}$. Let $G_{ab}$ be the subgraph of $G$ induced by these two colors and let $Q$ be the connected component of $G_{ab}$ containing the vertices colored with $b$ in $G_{2k-1}$. 
Since $\Gamma_{i-2} \subseteq \Gamma_{0}$ by Claim \ref{Clm:C1InC2i+1_C2iInC0} and $a \not \in \Gamma_{i-2}$, $Q$ does not contain any vertex in $\Gamma_{i-2}$. 
Hence $Q$ does not contain any vertex colored with $a$ in $G_{0}$ and $G_{ab}$ is not connected, a contradiction with Claim \ref{Clm:GabConnected1}.
\end{PrfClaim}

\begin{Clm}\label{Clm:vivi_1MaximumWeight}
For an odd index $i$ ( $1\leq i\leq 2k-1$), $v_{i-1}v_{i}$ is an edge with maximum weight.
\end{Clm}
\begin{PrfClaim} Since $\Gamma_{1} \subseteq \Gamma_{i}$ and $\Gamma_{i-1} \subseteq \Gamma_{0}$ by Claim \ref{Clm:C1InC2i+1_C2iInC0}, let us prove that $\Gamma_{0} - \Gamma_{i-1} \subseteq \Gamma_{i}$. 
Assume that some color $a \in \Gamma_{0} -\Gamma_{i-1}$ does not appear in $\Gamma_{i}$. Let $b$ be any color in $\Gamma_{2k-2}$ (recall that $\Gamma_{2k-2} \subseteq \Gamma_{0}$ by Claim \ref{Clm:v2k-1v2k-2MaximumWeight})
 and let $G_{ab}$ be the subgraph induced by these two colors. Since $a$ does not appear in $\Gamma_{i} \cup \Gamma_{i-1}$ but appears in $\Gamma_{2k-1}$ by Claim \ref{Clm:v2k-1v2k-2MaximumWeight}, 
the connected component $Q$ of $G_{ab}$ containing the vertices of color $a$ in $G_{2k-1}$ is distinct from the component containing the vertices of color $a$ in $G_{0}$.

Let us now exchange the colors $a$ and $b$ on $Q$. In this new coloring of $G$, let $Q^{'}$ be the connected component of the subgraph induced by the colors $a$ and $c$ where $c$ is any color in $\Gamma_{1}$. 
Since $a$ is always lacking in the sets of color $\Gamma_{i}$ as well as in $\Gamma_{i-1}$, $Q_{'}$ does not contain any vertex colored with $a$ in $G_{0}$. 
We can thus proceed to a new exchange of colors $a$ and $c$ on $Q_{'}$. The color \Enleve{$c$}\Ajoute{$a$} is now available to coloring $v$, a contradiction.

But now, since $\chi_i=|\Gamma_i|=|\Gamma_1|+|\Gamma_0|-|\Gamma_{i-1}|$ and $\chi_{i-1}=|\Gamma_{i-1}|$, we have

$\chi_i+\chi_{i-1}=\chi_0+\chi_1$, in other words $v_{i-1}v_{i}$ is an edge with maximum weight.
\end{PrfClaim}

\begin{Clm}\label{Clm:C2i-1DansC2i+1}
For any odd index $i$ ( $1\leq i\leq 2k-3$), $\Gamma_{i} \subseteq \Gamma_{i+2}$.
\end{Clm}

\begin{PrfClaim} Obvious by virtue of Claims \ref{Clm:vivi_1MaximumWeight} and \ref{Clm:C2iDansC2i-2}.
\end{PrfClaim}

\begin{Clm}\label{Clm:EveryComponentSize2}
For any index $i$ ($0 \leq i \leq 2k-1$), $G_{i}$ has at least two vertices
\end{Clm}
\begin{PrfClaim}
Assume to the contrary that $G_{i}$ is reduced to a single vertex for some $i\in\{0,\ldots 2k-1\}$.

If $i$ is even then, by Claim \ref{Clm:vivi_1MaximumWeight}, $v_iv_{i+1}$ has maximum weight and the unique vertex in $G_i$ has maximum degree in $G[V_i\cup V_{i+1}]$. Consequently, by Theorem \ref{Theorem:NeighborhoodEdgeMaximumWeight}, 
$G_{i-1}$ is reduced to a single vertex. But now, by Claim \ref{Clm:C2i-1DansC2i+1}, $\Gamma_1\subseteq \Gamma_{i-1}$, that means $\chi_1=1$ since $|\Gamma_{i-1}|=1$. 
By Claim \ref{Clm:C2iDansC2i-2}, $|\Gamma_{i+2}|=|\Gamma_{i+4}|=\ldots |\Gamma_{2k-2}|=1$. In addition, $v_0v_{2k}$ has maximum weight, it follows $|V_{2k-1}|=1$. 
Let us set $\Gamma_{2k-2}=\{a\}$ and $\Gamma_{2k-1}=\Gamma_1=\{b\}$,
 of course $a\in \Gamma_0$.\\
\indent We claim that $\Gamma_0=\{a\}$. Assume, on the contrary, that in $\Gamma_0$ there is a color, say $c$, distinct from $a$. The subgraph $G_{bc}$ induced by the vertices of $G$ colored with $b$ and $c$ is not connected since $c\notin \Gamma_{2k-2}$.
In this conditions, we could exchange the colors $b$ and $c$ on the component of $G_{bc}$ which contains vertices of $V_0$ and use the color $c$ for the coloring of the vertex $v$, a contradiction.\\
\indent Hence, $|\Gamma_0|=1=\chi_0$ and $\chi_0+\chi_1=2$. 
In other words for $0\leq i\leq 2k$, $V_i$ is a stable set and $G$ is an empty expansion of an odd hole, a contradiction (see \cite{AraKarSub11}).\\

\indent When $i$ is odd, the edge $v_{i}v_{i-1}$ having maximum weight in $\Gamma_{2k+1}$ by Claim \ref{Clm:vivi_1MaximumWeight}, $G_{i+1}$ is reduced to a single vertex 
by Theorem \ref{Theorem:NeighborhoodEdgeMaximumWeight} and the above reasoning holds.

\end{PrfClaim}

To end our proof assume first that $k\geq 3$. An edge $v_{i}v_{i-1}$ with $i$ odd being of maximum weight in $H$ by Claim \ref{Clm:vivi_1MaximumWeight}, one of $G_{i+1}$ or $G_{i-2}$ 
must be reduced to a single vertex by Theorem \ref{Theorem:NeighborhoodEdgeMaximumWeight}, a contradiction with Claim \ref{Clm:EveryComponentSize2}.

Hence from now on $k=2$. Let $G'=G[V_0\cup V_1]$. By Claim \ref{Clm:EveryComponentSize2}, $|V_{i}| \geq 2$ for $i=0 \ldots 4$. Moreover, $\Delta_1\geq 1$, otherwise the edge $v_0v_4$ would have maximum weight in $C_{2k+1}$ 
and $|V_3|=1$ by Theorem \ref{Theorem:NeighborhoodEdgeMaximumWeight}, a contradiction with Claim \ref{Clm:EveryComponentSize2}.

Assume that $|V_2|\geq |V_1|$ and let $w$ be a vertex of maximum degree in $G_1$. We have
$$\Delta\geq d(w)\geq|V_0|+|V_2|+\Delta_1\geq |V_0|+|V_2|+1\geq \Delta_0+|V_1|+2=\Delta(G')+2.$$

Consequently $\lceil\frac{\omega(G)+\Delta(G)+1}{2}\rceil\geq \lceil\frac{\omega(G')+\Delta(G')+1}{2}\rceil+1$ and by Theorem \ref{Theorem:Complement}, $\lceil\frac{\omega(G')+\Delta(G')+1}{2}\rceil+1\geq \chi_0+\chi_1$. Hence $\lceil\frac{\omega(G)+\Delta(G)+1}{2}\rceil\geq \chi_0+\chi_1+1$, 
a contradiction since $\chi_0+\chi_1+1$ is precisely the chromatic number of $G$.

Hence we must suppose that $|V_2| < |V_1|$. Since $v_2v_3$ is an edge of maximum weight in $C_{2k+1}$  with $v$ in the neighborhood of $G_{4}$ in the expansion, 
we could have chosen this edge as the edge $v_{0}v_{1}$. With the same reasoning we should obtain that $|V_1| < |V_2|$, a contradiction.
\end{Prf}

\begin{Thm}\label{Thm:OddExpansionComposanteBipartite}
If $G=C_{2k+1}(G_0\ldots G_{2k})$ is an expansion of an odd hole such that $G_{i}$ induces a bipartite graph for some $i\in\{0\ldots 2k\}$ then Conjecture \ref{Conj:Reed} holds for $G$.
\end{Thm}
\begin{Prf} Assume that $G$ is a counter-example to Conjecture \ref{Conj:Reed}. By Corollary \ref{Corollary:UpperBoundChromaticNumberExpansionOddHole}, $\chi(G) \leq \chi_{i}+\chi_{i+1}+1$ 
when $v_{i}v_{i+1}$ is an edge with maximum weight. When $\chi(G) =\chi_{i}+\chi_{i+1}$, we have a contradiction with Lemma \ref{Lemma:ReedPourSousGrapheQuiAtteintLeChromaticNumber}. 
When $\chi(G) = \chi_{i}+\chi_{i+1}+1$, one component of $G$ must be reduced to a single vertex by Corollary \ref{Cor:OtherNeighborhoodEdgeMaximumWeight}, 
a contradiction with Theorem \ref{Thm:OddExpansionComposanteSingleton}.
\end{Prf}
\begin{Thm}\label{Thm:OddExpansionComposanteEqualChromaticNumber}
If $G=C_{2k+1}(G_0\ldots G_{2k})$ is an expansion of an odd hole  such that $\chi_{i}=q \geq 1$  for all $i\in\{0\ldots 2k\}$ then  Conjecture \ref{Conj:Reed} holds for $G$.
\end{Thm}

\begin{Prf}
Assume to the contrary that $G$ is a counter-example to Conjecture \ref{Conj:Reed}. Since every edge of $H$ has maximum weight, for every $i\in\{0\ldots 2k\}$ $V_{i-2}$ or $V_{i+1}$ 
induces a complete graph on exactly $q$ vertices,  by the hypothesis and Theorem \ref{Theorem:NeighborhoodEdgeMaximumWeight}. Hence, it is not difficult to see that at least two components, 
say $V_{0}$ and $V_{1}$, are isomorphic to $K_{q}$. We have thus $\omega \geq 2q$ and $\Delta \geq 3q-1$ (a vertex in $V_{1}$ has $q$ neighbors in $V_{0}$, $q-1$ in $V_{1}$ and at least $q$ neighbors in $V_{2}$) which leads to
$$\lceil\frac{\omega(G)+\Delta(G)+1}{2}\rceil \geq \lceil\frac{5q}{2}\rceil.$$

By Theorem \ref{Theorem:ChromaticNumberExpansionOddHole} we have $\chi(G) \leq \lceil\frac{5q}{2}\rceil$, a contradiction.
\end{Prf}

\begin{Thm}\label{Thm:OddExpansionComposanteAtMost3}
If $G=C_{2k+1}(G_0\ldots G_{2k})$ is an expansion of an odd hole such that $\chi_{i} \leq 3$  for all $i\in\{0\ldots 2k\}$ then Conjecture \ref{Conj:Reed} holds for $G$.
\end{Thm}
\begin{Prf} Assume that $G$ is a counter-example to Conjecture \ref{Conj:Reed}.  If some component has chromatic number at most $2$, we have a contradiction with 
Theorem \ref{Thm:OddExpansionComposanteBipartite}. Hence we must suppose that each component has chromatic number $3$, a contradiction with 
Theorem \ref{Thm:OddExpansionComposanteEqualChromaticNumber}
\end{Prf}

The following lemma will be useful in the next theorem. Its proof is 
standard and left to the reader.
\begin{Lem} \label{Lemma:4chromaticComponent} Let $K$ be a graph with chromatic number $4$.
 \begin{itemize}
 \item if $K$ has $5$ vertices then $K$ contains a $K_{4}$ 
 \item if $\omega(K)=2$ then $K$ has at least $8$ vertices.
 \end{itemize}
\end{Lem}

\begin{Thm}\label{Thm:OddExpansionComposanteAtMost4}
If $G=C_{2k+1}(G_0\ldots G_{2k})$ is an expansion of an odd hole such that 
$\chi_{i} \leq 4$  for all $i\in\{0\ldots 2k\}$ then Conjecture \ref{Conj:Reed} 
holds for $G$.
\end{Thm}
\begin{Prf} Assume that $G$ is a  counter-example to Conjecture \ref{Conj:Reed}.  If some component has chromatic number at most $2$, we have 
a contradiction with Theorem \ref{Thm:OddExpansionComposanteBipartite}. 
Hence we must suppose that each component has chromatic number $3$ or 
$4$. If no component has chromatic number $4$, we have  a contradiction 
with Theorem \ref{Thm:OddExpansionComposanteEqualChromaticNumber} as 
well as if every component has chromatic number $4$. Hence we can 
suppose that at least one component has chromatic number $3$ and at 
least one component has chromatic number $4$. This forces immediately 
$\chi_{0}+\chi_{1}=7$ or $8$. Let us remark also that $\omega \geq 4$.

We have $\chi(G)=9$ or $\chi(G)=10$  and, 
obviously, $\lceil\frac{\omega(G)+\Delta(G)+1}{2}\rceil \geq 9$ as soon as 
$\omega(G) + \Delta(G) \geq 16$ and $\lceil\frac{\omega(G)+\Delta(G)+1}{2}\rceil 
\geq 10$ as soon as $\omega(G) + \Delta(G) \geq 18$.

\begin{Clm} \label{Claim:NoComponentWith8verticesAtLeast}
Every component has at most $7$ vertices
\end{Clm}
\begin{PrfClaim}
Assume to the contrary  that some component $V_{i}$ has at least $8$ 
vertices. If $\Delta_{i+1} \geq 3$  then $\Delta \geq 14$. Hence $\omega(G) 
+ \Delta(G) \geq 18$ and Reed's conjecture holds for $G$, a contradiction. 
If $\Delta_{i+1} \leq 2$ then $V_{i+1}$ must be isomorphic to a  a 
triangle by Brook's Theorem. We have thus $\omega(G) \geq 5$ and $\Delta(G)
\geq 13$ and Reed's conjecture holds for $G$, a contradiction.
\end{PrfClaim}

From now on, we can consider that any component has at most $7$ 
vertices and hence, by Lemma \ref{Lemma:4chromaticComponent},   any 
$4-$chromatic component contains a triangle.

\begin{Clm} \label{Claim:NoTwoComponents4ChromaticConsecutive}
No two components with chromatic number $4$ are consecutive
\end{Clm}
\begin{PrfClaim}
Assume to the contrary  that for two consecutive components, $V_{i}$ and 
$V_{i+1}$, are such that $\chi_{i}=4$ and $\chi_{i+1}=4$. If these two 
components are isomorphic to a $K_{4}$ then any vertex in these 
components has degree at least $10$. Since a maximum clique of $G$ in 
this case  has at least $8$ vertices, we have $\omega + \Delta \geq 18$.

If only one component is isomorphic to a $K_{4}$ (without loss of 
generality say that $V_{i}$ induces a $K_{4}$), then $\Delta_{i+1} \geq 
4$ by Brook's theorem and a vertex of maximum degree in $V_{i+1}$ has at 
least $11$ neighbors. Since a maximum clique of $G$ in this case  has at 
least $7$ vertices, we have $\omega + \Delta \geq 18$.

If no component is isomorphic to a $K_{4}$ then $\Delta_{i}$ and 
$\Delta_{i+1}$ are greater than $4$  by Brook's theorem. Moreover 
$V_{i}$  and $V_{i+1}$ contain at least $5$ vertices each. A vertex of 
maximum degree in $X_{i}$ has hence at least $12$ neighbors. Since a 
maximum clique of $G$ in this case  has at least $6$ vertices, we have 
$\omega + \Delta \geq 18$.

In each case we have a contradiction since $G$ satisfies Reed's conjecture.

\end{PrfClaim}

We can thus suppose that no two consecutive components have chromatic 
number $4$. In that case we can remark that $\chi(G)=9$. To end our proof, it is thus sufficient to show that 
$\omega(G)+\Delta(G) \geq 16$.

Without loss of generality, assume that $\chi_{0}=4$. By Claim 
\ref{Claim:NoTwoComponents4ChromaticConsecutive} we have $\chi_{2p}=3$ 
and $\chi_{1}=3$.

If $V_{0}$ induces a $K_{4}$ then either  $V_{2p}$ or $V_{1}$ contain a 
triangle and hence  $\omega \geq 7$  or have no triangle and $V_{2p}$ 
and $V_{1}$ contain at least $4$ vertices each. In the first case a 
vertex in $V_{0}$ has at least $9$ neighbors and $\omega(G)+\Delta(G) \geq 16$.
 In the second case we have $\omega(G) \geq 5$ and  a vertex in $V_{0}$ 
has at least $11$ neighbors. We get then $\omega(G)+\Delta(G) \geq 16$.

Assume now that  $V_{0}$ does not induce a $K_{4}$ then $\Delta_{0} \geq 
4$ by Brook's theorem. If $V_{2p}$ or $V_{1}$ contain a triangle then  
$\omega \geq 6$ and a vertex of maximum degree in $V_{0}$ has at least 
$10$ neighbors. We get then $\omega(G)+\Delta(G) \geq 16$.

If $V_{2p}$ and $V_{1}$ contain no triangle, these two sets must have at 
least $4$ vertices by Brook's theorem and a vertex of maximum degree in 
$V_{0}$ has at least $12$ vertices. Since $\omega \geq 5$ in that case, 
we get then $\omega(G)+\Delta(G) \geq 17$.

In each case we have a contradiction since $G$ satisfies Reed's conjecture.

\end{Prf}

Claim \ref{Claim:NoComponentWith8verticesAtLeast} in the proof of 
Theorem \ref{Thm:OddExpansionComposanteAtMost4} suggests that Reed's 
conjecture holds asymptotically for expansions of odd cycles.

\begin{Thm}\label{Theorem:AsympoticallyOddCyclesExpansion}
For every $k \geq 1$ and every $p \geq 1$, any expansion of an odd cycle $C_{2k+1}$  where each component has chromatic number at 
most $p$ and with at least $(2k+1)(5p-9)+1$ vertices satisfies Conjecture \ref{Conj:Reed}.
\end{Thm}

\begin{Prf} By Theorem \ref{Thm:OddExpansionComposanteAtMost4}, we can 
suppose that $p \geq 5$. Moreover, by Theorem 
\ref{Thm:OddExpansionComposanteBipartite}, we can suppose that each 
component has chromatic number at least $3$ and hence the maximum degree 
of each component must be at least $2$.  Let $G=C_{2k+1}(G_{0},G_{1} 
\ldots G_{2k})$ and assume that $\chi_{i} \leq p$ ($i=0 \ldots 2k$). By 
Corollary \ref{Corollary:UpperBoundChromaticNumberExpansionOddHole} we 
have $\chi(G) \leq \lceil \frac{5p}{2} \rceil$.

Suppose that some component $V_{i}$ ($i=0 \ldots 2k$) contains at least 
$5p-9$ vertices. Then a vertex in $V_{i+1}$ has degree at least $5p-4$. 
Since obviously  $\omega(G) \geq 4$ we have thus 
$\lceil\frac{\omega(G)+\Delta(G)+1}{2}\rceil \geq \lceil \frac{5p+1}{2} 
\rceil$. Hence $G$ satisfies Conjecture \ref{Conj:Reed} and the result follows.
\end{Prf}

\begin{Thm}\label{Theorem:C5_Expansion} If $G$ is a  $C_{5}$-expansion then Conjecture \ref{Conj:Reed} holds for $G$.
\end{Thm}
\begin{Prf}
Let $G=C_{5}(G_{0},G_{1},G_{2},G_{3},G_{4})$ and assume by
contradiction  that $G$ does not satisfy Conjecture \ref{Conj:Reed}.
Let $p=\min \chi(G_{i})\ i=0,
\ldots, 4$, by Theorem \ref{Thm:OddExpansionComposanteBipartite} we  have
$p \geq 3$.

We suppose that $\chi(G_{0})+\chi(G_{1})$ is maximum among the pairs
of consecutive components of $G$ and we denote $G'=G[V_0\cup V_1]$.  By Theorem
\ref{Theorem:NeighborhoodEdgeMaximumWeight},  $G_{4}$ or $G_{2}$ induce a
complete graph on $p$ vertices. We  assume  that
$G_{4}$ is this component and  there is a vertex in
$V_{0}$ whose degree in $G'$ is maximum. Moreover, Theorem \ref{Theorem:NeighborhoodEdgeMaximumWeight} implies that $\chi_{2}+\chi_{3}=\chi_{0}+\chi_{1}$.

By Corollary \ref{Corollary:ChromaticNumberExpansionOddHole_If_Couterexample} we have $\chi(G)=\chi_{0}+\chi_{1}+\lfloor\frac{p+1}{2}\rfloor$.

We claim now that $|V_{2}|<|V_{1}|$ or $G_{1}$ is isomorphic to a
$C_{2s+1}$ with $s\geq 2$ (and henceforth $p=3$). Assume to the
contrary that $|V_{2}|\geq |V_{1}|$. Let $w$ be a vertex of maximum degree in
$G_{1}$. By Theorem \ref{Theorem:Complement} we have
$\chi(G_{0})+\chi(G_{1}) \leq \lceil
\frac{\omega(G^{'})+\Delta(G^{'})+1}{2} \rfloor$. Since $d(w) =
\Delta(G_{1})+|V_{0}|+|V_{2}| \geq \Delta(G^{'})+\Delta(G_{1})+1$ we
have $\Delta(G) \geq \Delta(G^{'})+\Delta(G_{1})+1$. By Brook's Theorem \cite{Bro41} we have
$\chi(G_{1}) \leq \Delta(G_{1})$ or $G_{1}$ is an odd chordless
cycle. When $\chi(G_{1}) \leq \Delta(G_{1})$, we get

\begin{equation}\label{Equation:1}
   \lceil \frac{\omega(G)+\Delta(G)+1}{2} \rceil \geq  \lceil \frac{\omega(G^{'})+\Delta(G^{'})+p+1+1}{2}\rceil.
\end{equation}
Which leads to $\lceil \frac{\omega(G)+\Delta(G)+1}{2} \rceil \geq\chi(G_{0})+\chi(G_{1})+ \lfloor\frac{p+1}{2}\rfloor = \chi(G)$, a contradiction.

If $G_{1}$ is isomorphic to a $C_{2s+1}$ with $s\geq 2$ we have $\omega(G^{'})=\omega(G_{0})+2$, $\omega(G)\geq \omega(G_{0})+3$ and  $\Delta(G) \geq \Delta(G^{'})+3$. Hence
\begin{equation}\label{Equation:10}
    \lceil \frac{\omega(G)+\Delta(G)+1}{2} \rfloor \geq  \lceil \frac{\omega(G^{'})+1+\Delta(G^{'})+3+1}{2}\rfloor.
\end{equation}
Which leads to $\lceil \frac{\omega(G)+\Delta(G)+1}{2} \rfloor \geq\chi(G_{0})+\chi(G_{1})+ 2 \geq \chi(G)$, a contradiction.

If $G[V_{2}\cup V_{3}]$ contains a vertex of maximum degree in $V_{2}$, by Theorem \ref{Theorem:NeighborhoodEdgeMaximumWeight}, $G_{1}$ is  a complete graph on $p$ vertices, a contradiction with $|V_{2}|<|V_{1}|$. Hence a vertex of maximum degree in $G[V_{2}\cup V_{3}]$ must be a vertex of $G_{3}$. By application of the above technique we can thus prove that $|V_{1}|<|V_{2}|$ or $G_{2}$ is isomorphic to a
$C_{2s+1}$ with $s\geq 2$. In the first case, we get a contradiction with  $|V_{2}|<|V_{1}|$. In the latter case, we can conclude as above. 

\end{Prf}
\begin{Thm}\label{Thm:Reed_à_2_Près}
If $G=C_{2k+1}(G_0\ldots G_{2k})$ is an expansion of an odd hole then $\chi(G)\leq\lceil\frac{\omega(G)+\Delta(G)+1}{2}\rceil$+1.
\end{Thm}
\begin{Prf}
 We consider an optimal colouring of $G$. Let us denote $\displaystyle{p=\min_{0\leq i \leq 2k} \chi_{i}}$. 

If $p$ is even we have $\chi(G)\leq\lceil\frac{\omega(G)+\Delta(G)+1}{2}\rceil$ (Corollary \ref{Corollary:ReedQuandMinChromaticEstPair}). 
Consequently, in the following, we suppose that $p$ is odd. 

Let $j\in\{0,\ldots 2k\}$ such that $\chi_{j}=p$. We choose some colour used for the colouring of $G_j$, say $c_j$ and we denote $S_j$ as the set of vertices of $G_j$ being coloured with $c_j$.

We set $G'_j=G[V_j-S_j]$ and for $i\neq j$ we set $G'_i=G_i$. 

$G'=C_{2k+1}(G'_0,\ldots G'_{2k})$ is an odd expansion such that the minimum chromatic number of its components is $p-1$. Since $p-1$ is even, again by Corollary \ref{Corollary:ReedQuandMinChromaticEstPair}, we have
$\chi(G')\leq \lceil\frac{\omega(G')+\Delta(G')+1}{2}\rceil$  and consequently $\chi(G')\leq \lceil\frac{\omega(G)+\Delta(G)+1}{2}\rceil$.

But now, given an optimal colouring of $G'$, we can obtain an optimal colouring of $G$ with only one additionnal colour (for the vertices of $S_j$). In other words, $\chi(G)\leq \chi(G')+1$. The result follows.

\end{Prf}
In a further paper \cite{FouVan2012a}, we will use the above results in order to extend  a number of the results given in \cite{AraKarSub11}.

\bibliographystyle{plain}
\bibliography{BibliographieReed}

\end{document}